\documentclass[11pt,twoside]{article}
\usepackage{asp2004}
\usepackage{psfig}
\usepackage{epsf}
\usepackage{graphics}
\usepackage{lscape}
\def\edcomment#1{\iffalse\marginpar{\raggedright\sl#1\/}\else\relax\fi}
\reversemarginpar

\begin{document}
\title{Triggering the Formation of Massive Clusters}
\author{Bruce G. Elmegreen}
\affil{IBM Research Division, T.J. Watson Research Center,
Yorktown Hts., NY 10598 USA bge@watson.ibm.com}

\begin{abstract}
There are at least 2 distinct mechanisms for the formation of
young massive clusters (YMC), all of which require galactic-scale
processes. One operates in harrassed fragile galaxies, in the
dense cores of low mass galaxies, at the ends of spiral arms, or
in galactic tidal shocks where transient and peculiar high
pressures make massive clouds at high densities. The result of
this process is usually only one or two YMC without the usual
morphologies of local star formation, i.e., without hierarchical
structure and a continuous power law distribution of cluster
masses up to the largest mass.  The other operates in the more
usual way: continuously for long periods of time in large parts of
the interstellar medium where the ambient pressure is already high
as a result of the deep potential well from background stars and
other gas. This second process makes clusters in a hierarchical
fashion with size-of-sample effects, and tends to occur in nuclear
rings, merger remnants, and even the ambient ISM of normal
galaxies if the star formation rate is high enough to sample out
to the YMC range.
\end{abstract}
\thispagestyle{plain}

\noindent In Formation and Evolution of Massive Young Star
Clusters, Cancun, Mexico, November 17-21, 2003, eds.  Henny
Lamers, Linda Smith and Antonella Nota, Astronomical Society of
the Pacific (PASP Conference series), in press

\section{Triggering Low Mass Clusters}

Local clusters often show the signatures of high-pressure
triggering: they form in the heads of pressure-swept, cometary
clouds that are adjacent to older massive stars, or they form in
compressed layers or shells between expanding HII regions or wind
bubbles and the surrounding gas. Examples of the former include
Orion (Bally et al. 1987; Lada, et al. 1991; Reipurth, Rodriguez
\& Chini 1999), the Eagle nebula (Hester et al. 1996), the rho Oph
core (de Geus 1992), and many places in the clouds surrounding 30
Dor in the LMC (Walborn et al. 2002). Examples of the latter
include the Carina nebula (Brooks, et al. 1998, 2001), the Rosette
nebula (Phelps \& Lada 1997), and other regions near 30 Dor. A
list of likely triggered regions is in Elmegreen (1998).

Sometimes clusters form at the tips of elongated clouds with no
obvious pressure source nearby. IC 5146 looks like this (Lada,
Alves \& Lada 1999): it is a long, filamentary cloud with most of
the star formation near the eastern tip. Another example is in the
Taurus region where most of the famous filaments have their star
formation toward the east (Elmegreen 2002), often with short-lived
molecules at these places, suggesting recent compression
(Hartquist et al. 2001). In these places, the star formation is
occurring at the most vulnerable places in the cloud where stray
pressure bursts would have the greatest cross section for
interaction. Even if these pressure sources cannot be identified
yet, perhaps because they were stray supernova whose other signs
have long disappeared, the peculiar positions of star formation
suggest some type of triggering was involved. In the case of
Taurus, the pressure seems to have come from the Orion OB
association.

\section{Triggering by Random Pressure Bursts}

The probability that random pressure excursions trigger star
formation is pretty low, especially for dense clusters of moderate
mass.  This conclusion comes from the probability distribution
function for pressure excursions of a certain magnitude. If we
consider a pressure source like an HII region, supernova or
wind-swept bubble, the radius increases with time as $R(t)$. This
function corresponds uniquely to a pressure dependence $P(t)$ for
each type of source, and thus there is a relation between volume
and pressure: $V(P)$. If these pressure bursts occur at a constant
rate, then the number density of small regions having a certain
pressure is proportional to the inverse of the pressure
derivative: $n(P)\propto1/\left(dP/dt\right)$. This comes from the
one-to-one correspondence between pressure and time, and from the
resulting equality between the number distribution of pressure
events $n(P)dP$ between $P$ and $P+dP$, and the time distribution
of pressure, $P(t)dt$ between $t$ and $t+dt$ for constant rate
$P(t)$. This number density $n(P)$ combines with the volume
function, $V(P)$, to give the filling factor of regions with
pressure $P$: $f(P)=n(P)V(P)$ in linear intervals $dP$.

For HII regions, $f(P)\propto P^{-4.17}$, for bubbles and
supershells, $f(P)\propto P^{-4.5}$, and for the pressure-driven
snowplow phase of supernovae, $f(P)\propto P^{-5.2}$. These
relations come from the usual expansion laws for these region,
$R(t)$ (e.g., Weaver, et al. 1977; Cioffi et al. 1988).  The
summed contributions from these relations preserve the approximate
power-law form, $f(P)\simeq AP^{-4.5}$ or so, for constant of
proportionality $A$. If the summed filling factor from all
expansions is unity, then $1\sim\int AP^{-4.5}dP$, giving an
average ISM pressure $P_{ave}=1.4P_{min}$ for minimum pressure
$P_{min}$, and $f(P)=1.15\left(P/P_{ave}\right)^{-4.5}/P_{ave}$.
From this result we can determine the probability that the
pressure is between $P$ and $P+dP$, measured as the volume filling
factor, $f(P)dP$ for pressure in this range. Thus the probability
the pressure exceeds 10 times the average is
$0.31\left(0.1\right)^{3.5}\sim10^{-4}$, and the probability the
pressure exceeds 2 times the average is
$0.31\left(0.5\right)^{3.5}\sim0.03$.  These are very small
probabilities because the pressure from a power source decreases
rapidly during the expansion to larger volumes.  As a result,
significant {\it random} pressure bursts from HII regions,
supernovae, windy bubbles, and supershells are generally weak, on
the order of a factor of 2 or less, although they may be frequent.
This means that {\it pressure triggering from specific sources is
usually very localized -- in the same cloud complex, especially if
what is triggered is a dense star cluster at a typically high
pressure.}

A similar result may be gleaned from figure 3 in Kim, Balsara, \&
MacLow (2001), which shows the probability distribution function
for pressure in a turbulent medium.  This function is
approximately log-normal with a range of a factor of $\sim2$ in
either direction about the dominant pressure and a small plateau
at $10\times$ the dominant pressure from the young supernova
remnants.  Again it is clear that large pressure excursions around
the average ISM pressure should be rare.

\section{Energy requirements for triggering YMCs}

Another property of pressurized triggering can be inferred from
the energy requirements.  For pressure to be important in cloud
formation, the energy to move the gas has to be on the order of
$E\sim PV\sim Mv^2$ for ambient pressure $P$, moved volume $V$,
moved mass $M$, and ambient velocity dispersion $v$. With $v\sim7$
km s$^{-1}$, the formation of a YMC from a $10^6$ M$_\odot$ cloud
requires $10^{51}$ erg of energy during cloud formation. Moreover,
this energy has to be convergent so that ambient gas is compressed
into a cloud, not divergent like an explosion. We see from this
that only large clusters can trigger other large clusters, and
again that stray explosions are unlikely triggering sources in
main galaxy disks, primarily because of their divergent nature.
This means that an isolated YMC in a main galaxy disk was probably
not triggered by compression from a stray explosion of the most
common type. Instead we should look for a history of globally
convergent flows, such as kpc-scale instabilities, galaxy-wide
turbulence, or galaxy interactions. In the absence of such
converging flows, massive clusters need high densities and
pressures in the {\it ambient} medium, as is often the case in
galactic nuclei.  An example of nuclear triggering in a BCD galaxy
could be Markarian 86 (Gil de Paz et al. 2000; 2002).  Several
other possible examples are in Saito, Kamaya, \& Tomita  (2000),
and Cair\'os et al. (2001).

\section{Large-Scale converging flows that do not seem to trigger YMCs}

Galactic scale instabilities easily make $10^6-10^7$ $M_\odot$
clouds and these could in principle make YMCs but in the main
disks of galaxies these regions are generally at the ambient
pressure, which is low, and so their average densities are also
low. That is, the stars they make are rarely collected into a
$\sim5$ pc size YMC but they are dispersed throughout a kpc
region, like Gould's Belt, with a hierarchical pattern and dense
clusters only on the smallest scales. Most clusters within several
kpc of the Sun along with their associated ``giant'' molecular
clouds are like this: they are only small pieces in much larger
cloud complexes that dot the spiral arms with $\sim2-3$ kpc
separations (e.g., see review in Elmegreen 2002). Often the
clusters themselves look triggered inside these clouds, on
$\sim10$ pc scales.  For example, the outer Galaxy survey by Heyer
et al. (2001) shows two large complexes, one associated with the
HII regions W3/4/5 and another associated with NGC 7538. Inside
these large regions, which are separated by $\sim25^\circ\sim1$
kpc of relatively low CO emission, there are clusters and
molecular clouds, many of which look triggered or perturbed by
high pressure events.  The IRAS point sources in the W3/4/5
region, for example (Carpenter, Heyer \& Snell 2000), are mostly
at the tips of cometary CO structures.  No YMC's have formed
although the total molecular masses in these two regions are large
enough.

Clearly, galactic-scale triggering does not necessarily mean the
formation of a YMC. Most galactic disk star formation processes
are massive because their length scales are big, but they are not
nearly high enough in pressure to make a YMC. The pressure inside
a cloud core where a cluster forms is $\sim 0.1GM^2/R^4$, i.e.,
proportional to the squared column density. The density of a dense
star cluster is typically $10^3-10^4$ M$_\odot$ pc$^{-3}$, so the
column density is this multiplied by the radius, or
$\sim10^2-10^4$ M$_\odot$ pc$^{-2}$, with the upper end typical of
YMCs. These column densities translate into pressures equal to
$0.1G$ times the square of their values, which are in the range
$10^{-10}$ to $10^{-6}$ erg cm$^{-3}$, or $10^{6}$ to $10^{10}$
times Boltzman's constant, $k_B$. The ambient pressure in a
typical disk is only $10^4k_B$. Thus, massive high-pressure clouds
need {\it highly} compressive, galactic scale events, not mildly
compressive galactic-scale events.  Instabilities involved with
the formation of spiral arms and their regularly-spaced giant
clouds are not usually dense enough to make dense massive clusters
by themselves. Further collapse into massive dense cores, possibly
in combination with pressurized triggering by very energetic older
clusters, are required in addition.

\section{Galactic processes that may have triggered YMCs}
\label{sect:x}
\subsection{Main spiral disks}

There are several examples of galactic processes that seem to have
triggered the formation of a YMC. In NGC 6946, a YMC with $10^6$
M$_\odot$ of stars lies at the end of a spiral arm, suggesting an
asymmetric and perhaps unusually strong collapse of gas by an
unbalanced gravitational force (Elmegreen et al. 2000; Larsen et
al. 2002).  Remarkably, another YMC in the same galaxy is at the
tip of a different arm (S. Larsen, this conference), reinforcing
this idea. Inside the first region, the history of star formation
suggests a period lasting $\sim40$ My producing distributed small
clusters with an interruption of this mode $\sim15$ My ago during
which the YMC was the primary star-forming event.  At $\sim5$ My
ago, the distributed star formation began again, lasting until
today. Perhaps the imbalanced collapse made a massive dense cloud
and the first generation of stars in this cloud compacted the
remainder to make the YMC at extreme pressures (Larsen et al.
2002). Likely remnants of this cloud are still visible. The 30 Dor
cluster in the LMC has a similar two-step structure with
distributed, slightly older stars and clusters along the
periphery, many of which seemed to have formed there, and a
compact younger cluster in the center (Selman et al. 1999).

Although this outside-in morphology suggests convergent
triggering, another possibility is that the dense cluster is a
remnant of a prolonged coalescence of many small clusters that
formed in a distributed fashion throughout the region.  The age of
the YMC in NGC 6946 has the average value of the ages of the
smaller clusters around it, and the collision time works out for
this model (Elmegreen et al. 2000).  But in 30 Dor, the central
cluster seems younger than the average of the other clusters, and
in that case, the coalescence model does not work.

The 30 Dor region may have suffered from a large-scale compression
resulting from its motion through the galactic halo (de Boer et
al. 1998). This compression would have pushed on the whole eastern
side of the LMC and made a giant CO cloud that extends for $\sim1$
kpc south of 30 Dor, with 30 Dor at its northern tip.

Other small galaxies with YMCs look similarly perturbed (Billett,
Hunter, \& Elmegreen 2002). NGC 1569 is a classic example of a
small galaxy with YMC's (Waller 1991; Ho \& Filippenko 1996a;
Hunter et al. 2000) and it has a peculiar stream of HI extending
from far outside the optical radius to the very point in the disk
where the 2 YMC's are located (Waller 1991). It looks like this HI
crashed into the disk and compressed the ambient gas to make the
clusters out of the resulting massive dense clouds.

Large galaxies can be perturbed by collisions also. The NGC
2207/IC 2163 pair suffered a grazing collision with a
peri-galacticon $\sim40$ My ago. One of these galaxies was
perturbed in a prograde, in-plane sense, and its outer disk
responded by falling inward for a half-epicycle and crashing
against other disk material that did not fall in as quickly. As a
result, a galactic-scale shock front formed having the overall
shape of an pointed-oval or eye (Elmegreen et al. 1991). This
shock front contains several YMC (Elmegreen et al. 2001).

\subsection{Galactic nuclei}

The centers of some low-mass galaxies have YMCs, as does the
nuclear region of the Milky Way.  NGC 4214 is a dwarf galaxy with
a 4-5 My old YMC in the nuclear region (Leitherer et al 1996;
Billett et al 2002). Other examples are NGC 1705 (Meurer et al.\
1992; O'Connell et al.\ 1994; Ho \& Filippenko 1996b), and the
probable embedded clusters in NGC 5253 (Turner, Ho, \& Beck 1998),
He 2-10 (Conti \& Vacca 1994; Kobulnicky \& Johnson 1999), and NGC
2366 (Drissen et al.\ 2000).

The processes of collecting massive amounts of gas into dense
nuclear clusters are not known. They could form by spontaneous
gravitational instabilities in dense nuclear gas that is drawn in
from the outer disk by asymmetric forces and viscous accretion.

\subsection{Morphology and size of sample effects in YMC-triggered regions}

Some of these examples of YMC triggering do not seem to satisfy
the size-of-sample effects expected for a random ensemble of
clusters. What is observed is that the YMC is much more massive
than any other cluster in the galaxy or region. For the
size-of-sample effect to apply, the YMC would have to be the most
massive member of a continuous power law distribution of mass.
Perhaps our impression that the size-of-sample effect does not
apply in these cases is statistically insignificant because the
number of violations like this is very small. For example, in the
case of dwarf galaxies with unusually large YMCs, such as NGC
1569, it could be that the whole galaxy should be viewed as a
member of the ensemble, along with other whole galaxies, and not
just the individual clusters in one galaxy.  Then, if we were to
sample among many galaxies, we might expect that all of the
clusters in the composite would exhibit a smooth power law mass
distribution even if each galaxy alone has large deviations from
this.

Nevertheless, the observations give the impression that some
regions selectively produce YMC without making a proportional
number of low mass clusters.  A cluster mass distribution function
in these regions gives a slope that is significantly flatter than
$-2$ (for linear intervals of mass).  This is the case in IC 2163
and NGC 2207 mentioned above, where the slopes are $1.85\pm0.05$
and $1.58\pm0.12$ (Elmegreen et al. 2001). The most massive two
YMCs in IC 2163 are in the galactic tidal shock, and they have a
mass comparable to the most massive YMCs in the antennae, which
contains $\sim10\times$ more massive clusters overall.  Thus the
YMCs in IC 2163 do not appear to satisfy the normal size-of-sample
relation given the small number of other clusters in this galaxy.

Clusters in these regions also look odd because they are not part
of a hierarchical network of star formation consisting of small
clusters or associations inside larger star complexes.  This
morphological issue is related to the size-of-sample problem. The
YMCs are so large when the size-of-sample effect is violated that
they completely dominate the region without any significant
superstructure or substructure, as in a hierarchy.  An analogy
might be with Gould's Belt. A YMC can have all of the mass of
stars in the local Gould's Belt put into a cluster as dense as the
Trapezium cluster.  As it is, the Trapezium cluster and others
like it are a small part of the local hierarchy that has Gould's
Belt on the largest scale, divided into OB associations, OB
subgroups, and individual clusters on smaller scales.

\subsection{Summary: YMC triggering by galactic scale flows}

YMCs have been found in {\it peculiar, high-pressure} regions
including: (1) fragile galaxies like dwarf Irregulars undergoing
interactions that produce relatively major disturbances; (2)
short-lived tidal arms or caustics in the main disks of
interacting galaxies; (3) the leading surface of the LMC that may
be subject to ram pressure from motion through the Milky Way, and
(4) gas collection points in the centers of small galaxies.

These regions tend to be small and form few clusters overall, so
the YMCs do not appear to be accompanied by the usual power law
distribution of numerous low mass clusters.

Proposed cluster formation processes in these regions include: (1)
shock compression in a large part of a galaxy; (2) local
compression and collapse from extragalactic cloud impacts, strong
gravitational instabilities at the end of a spiral-arm, or
colliding supershells (Chernin, Efremov, \& Voinovich 1995); (3)
coagulation of smaller clusters, or accretion to the galactic
nucleus where long-term collection can produce a massive central
concentration of gas at high pressure.

\section{YMC formation in disk regions with high ambient pressure.}

The same processes of star formation that make small clusters in
the solar neighborhood can make YMCs if the ambient pressure is
high.  Then massive self-gravitating cloud cores will have
extremely large column densities and star formation in them can
produce a massive cluster.  These processes include turbulent
fragmentation, sequential triggering, and spontaneous
gravitational collapse.

High ambient pressures occur in nuclear rings, nuclear disks, and
merger remnants.  Giant clusters form as part of a statistical
ensemble of clusters with a near-universal mass function,
$n(M)dM=AM^{-2}dM$ for constant $A$ that depends on the star
formation rate (Elmegreen \& Efremov 1997). The YMCs also tend to
form in a hierarchical fashion (Zhang, Fall, \& Whitmore 2001).

In this mode of cluster formation, there are several
size-of-sample effects.  First, the maximum cluster mass increases
with the number of clusters (Whitmore 2003; Larsen 2002). This
comes from setting to unity the integral over the cluster
distribution function above the maximum mass:
$1=\int_{M_{max}}^\infty n(M)dM$, which gives $A=M_{max}$. Then
the total number of clusters is $\int_{M_{min}}^\infty
n(M)dM\simeq M_{max}/M_{min}$. It is seen that the number of
clusters counted down to some conventional minimum mass scales
with the maximum mass.  If the cluster mass function were a
slightly different power law, then this size of sample scaling
would be slightly different too.

A second size-of-sample effect is that the maximum mass of all the
clusters in a logarithmic interval of age increases linearly with
the age. This is because the number of clusters that ever formed,
plotted in equal logarithmic intervals of age, increases linearly
with age.  In fact, low mass clusters often cannot be seen at
great age because they are too faint.  But this correlation
between maximum mass and age does not need to correct for these
missing clusters. Many of the largest bound clusters are not
likely to disperse until a very old age -- older than the disk.
This method has been used by Hunter et al. (2003) to derive the
cluster mass function in the LMC.

The YMC formation rate may scale with the star formation rate per
unit area (Larsen \& Richtler 2000) by another size-of-sample
effect (Billett et al. 2002). The average star formation rate per
unit area scales with the gas column density to the power
$\sim1.4$ (Kennicutt 1998), and the gas pressure scales
approximately as the column density squared. Thus the gas pressure
scales as the star formation rate per unit area to the power 1.4.
At a given density $n$ that defines a cluster, the cluster mass
scales with the pressure as
\begin{equation}M\sim6\times10^3\;M_\odot\left(P_{int}/10^8\;{\rm
K\;cm^{-3}}\right)^{3/2} \left(n/10^5\;{\rm
cm}^{-3}\right)^{-2}.\label{eq:mass}\end{equation} The
normalization for this relation applies to the molecular core near
the Trapezium cluster in Orion (Lada, Evans \& Falgarone 1997). If
$M_{max}\propto P^{3/2}$ also, then $M_{max}\propto$ the star
formation rate per unit area through its similar pressure
dependence: both $M_{max}$ and SFR/Area $\propto P^{1.4}$ or
$P^{1.5}$. This is the maximum mass that can form in a region with
a certain pressure.

Such a mass is likely to form if a sufficiently large number of
clusters forms that $M_{max}$ is sampled in the ensemble.
Considering also the size-of-sample count given above,
$N=M_{max}/M_{min}$, we have two definitions for $M_{max}$: One
comes from the ISM pressure, which is related to the star
formation rate {\it per unit area}, and the other comes from the
total number of clusters through the product $NM_{min}$, which
depends on the total star formation rate (not per unit area). In a
galaxy with small area and a large pressure (a dwarf starburst)
the first M$_{max}$ can exceed the second. In this case, only a
fraction of these galaxies with high enough pressure to form a YMC
will actually do so because of limitations from the small size of
the sample.  In a very large galaxy with a low total star
formation rate, the second $M_{max}$ can dominate the first. Then
the maximum cluster mass should be small (because high pressure
regions are very rare) compared to the maximum mass expected from
the large number of clusters. In this second case, the cluster
mass function may end abruptly at a low value of $M_{max}$
(determined by the low pressure), and there should be
significantly more clusters than just one at this maximum mass.
There are no observations yet of this second size-of-sample effect
yet.

Most normal galaxies have a value of $M_{max}$ from pressure
limitations that is about the same as the value from the
size-of-sample effect. That is, there is typically a smooth power
law of cluster masses up to the one largest cluster.  (Exceptions
to this were discussed in Sect. \ref{sect:x}).  Then the number of
YMCs, or the fraction of the uv light in the form of YMCs,
increases approximately linearly with the star formation rate. For
a sample of galaxies that are all about the same size, as in the
Larsen \& Richtler (2000) sample (Billett et al. 2002), the
fraction of uv light in the form of YMCs also increases in direct
proportion to the star formation rate per unit area.

\section{On the origin of the cluster mass function}

\begin{figure}
\plottwo{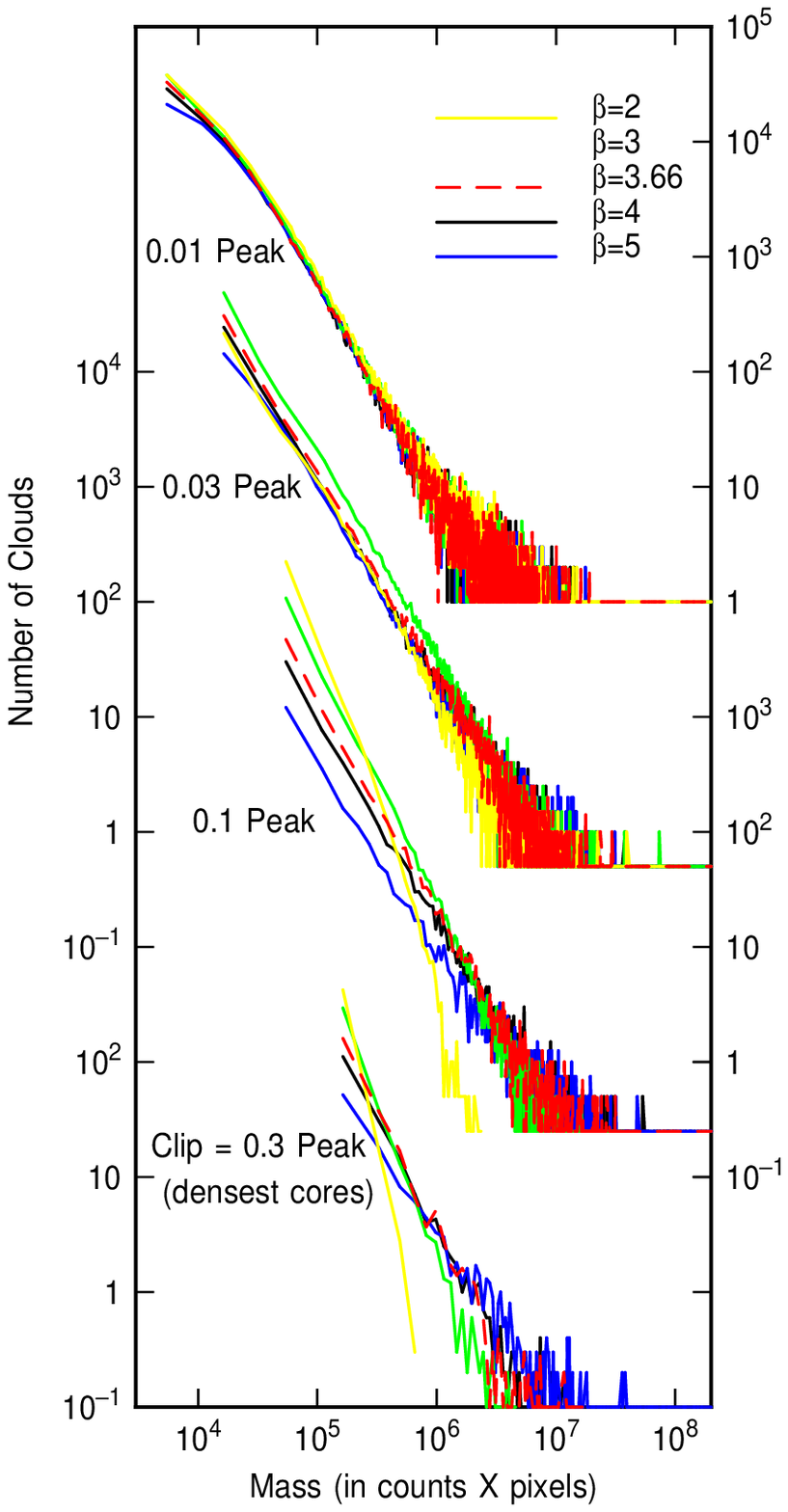}{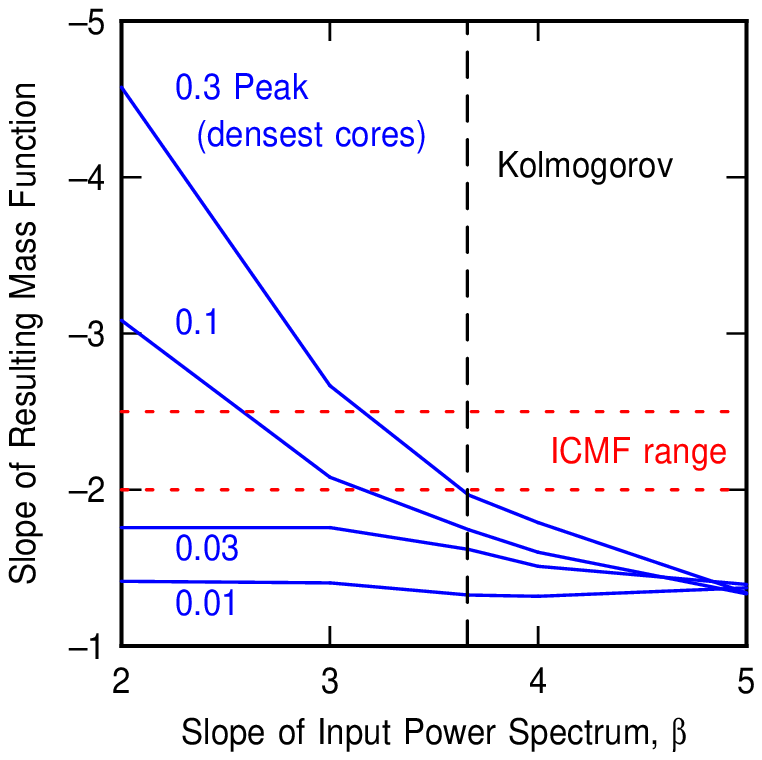} \caption{Models
of the cluster mass function made from cloud counting in Brownian
motion fractals. The thresholds for considering which density
maxima are clouds are given in terms of the peak density. The
slope of the power law distribution of $k^2$ in wavenumber space
is $-\beta$. The densest peaks, which presumably correspond to the
formation of clusters, are for the ``Clip=0.3 Peak'' case. The
mass function slope for these clouds is approximately $M^{-2}dM$
when the power spectrum has a power law corresponding to
Kolmogorov turbulence ($\beta=3.66$).  Note how mass spectra are
shallower for lower clipping levels, suggesting an origin for the
molecular cloud mass spectrum too.  This spectrum is shallower
than the cluster mass spectrum although both presumably come from
the same basic structure of turbulent gas.}\end{figure}

The characteristic $M^{-2}dM$ mass function of clusters could
follow from the hierarchical distribution of gas in a turbulent
medium. The densest regions of a random fractal have approximately
this mass function (Elmegreen 2002).  Figure 1 shows the mass
functions of distinct density maxima in a fractal Brownian motion
distribution of density that was made in the following way. First
a 3D grid in wavenumber-space, $(k_x,k_y,k_z)$, was filled with
random complex numbers having real and imaginary values between 0
and 1. This noise was multiplied by a power $-\beta/2$ of the
distance to the origin $k=\left(k_x^2+k_y^2+k_z^2\right)^{1/2}$.
The inverse Fourier transform of this noise cube was taken, giving
a cube twice as large in each dimension filled with positive and
negative real numbers having a Gaussian distribution.  This is a
Brownian motion fractal. To simulate the density distribution in
turbulence better, another cube is made from the exponential of
these positive and negative real values. After this, the numbers
are all positive and they have a log-normal distribution, as does
the density field in isothermal turbulence (Ostriker, Gammie \&
Stone 1999; Padoan et al. 2000; Klessen 2000; Ostriker et al.
2001; Li et al. 2003). A clump-finding algorithm was applied to
this model density field to get mass spectra. The mass spectra
depend on the minimum density that is accepted for a cloud. If the
minimum density is very high, then only the densest regions are
counted. The result would be most representative of star clusters,
which form in the densest part of the turbulent ISM. If the
minimum density is low, then the mass function should be more
representative of molecular clouds, or perhaps diffuse clouds,
which have lower densities compared to the peak.

On the left in figure 1 four sets of mass distributions are shown,
one for each cut-off density that defines the acceptable clouds.
The different curves for each are the mass functions for different
powers $\beta$. A Kolmogorov velocity field would have a power
$\beta=11/3$, which is one of the curves in the figure assuming
the same power law for density.  Clearly the mass functions are
all approximately power laws. Noise at the high mass end where
there are only a few clusters prevents an extrapolation there. The
power laws steepen as $\beta$ decreases. A low $\beta$ fractal has
a lot of structure on large wavenumbers, and this means there are
proportionally more low mass clumps, giving a steeper mass
function.

The right hand side of figure 1 shows the slopes as a function of
$\beta$ for 4 different density cutoff values. The vertical dashed
line is the power spectrum for a Kolmogorov velocity field. The
range of observed slopes for the initial cluster mass function
(ICMF) is shown. In the model, the slope of the mass spectra of
the highest density clumps, which are those with a cutoff density
equal to 0.3 times the peak density in the whole fractal, is equal
to the observed value of $-2$ when the power spectrum is
Kolmogorov.

This example suggests that clusters that form in the dense regions
of a turbulent ISM will have a mass function close to $M^{-2}dM$
as a result of turbulent fragmentation. This type of function is
also expected from simpler arguments.  For a scale-free density
distribution with a hierarchical nature, each clump contains
subclumps. If the number of subclumps per clump is constant, then
the entire mass of all the gas is present in the summed masses of
the clumps on each scale. That is, the smallest clumps contain all
the mass, but this same gas is also what comprises the
next-smallest clumps, and so on. In this case, the total mass in
each equal logarithmic interval of mass is constant, so
$M\xi(M)d\log M=$ constant, which means $M\xi(M)=$ constant.
Converting the mass function in log intervals, $\xi(M)$, to the
mass function in linear intervals, $n(M)$, by writing $\xi(M)d\log
M=n(M)dM$, we get $n(M)\propto\xi(M)/M$ and therefore $n(M)\propto
1/M^2$, as observed in figure 1.

\section{Conclusions}

Sequential triggering is common in star forming regions. Yamaguchi
et al. (1999, 2001a,b) estimate that 10\%-50\% of star formation
in the LMC and Milky Way is triggered by HII regions. However, the
triggered regions in the main disks of galaxies are usually too
small to make YMCs. Triggering by these same mechanisms but in the
nuclear regions of galaxies could make YMCs because the ambient
density and pressure are higher there than in the main disks. Or,
if there are a very large number of star-forming regions, some few
might achieve the high masses and densities required even in the
main disks.

Random triggering by stray supernovae or other pressure bursts
occurs outside star-forming regions too, but this process is rare
and the pressures are usually too weak and too divergent to make
YMCs (again, this statement is limited to the main disks of spiral
galaxies).

Galaxy-scale processes commonly move around the required gas mass
to make a YMC, but these processes are generally too low in
pressure to get the required high densities.

The observations of YMCs suggest two triggering mechanisms. One
applies to peculiar one-time events and special places that have
extraordinarily high energy inputs. These appear to make YMCs
without the usual $M^{-2}dM$ power law distribution of lower mass
clusters, i.e., to make YMCs almost exclusively. This conclusion
is uncertain because the sampling statistics are poor at the
present time. Examples include low mass galaxies which, because of
their low rotation and rms motions, are loosely bound and somewhat
fragile, making them easily perturbed by other small galaxies or
intergalactic gas clouds (e.g., Taylor 1997). The dense cores of
low mass galaxies also make YMCs, perhaps because of steady
accretion from the outer disk and high pressure from self-gravity.
The collapsed tips of spiral arms have produced YMCs in two cases,
suggesting these regions are more catastrophically unstable than
spiral arm midpoints. Tidal shocks in interacting galaxies can
make conditions right for YMC formation too by compressing large
parts of the ISM for a short time.

Another process that can make YMCs without a proportional number
of small clusters is cluster coalescence. Clusters of modest mass
that are born with the normal power law distribution of smaller
clusters around them may accrete a high fraction of these clusters
and become a YMC with few lower mass clusters remaining.

There is also a second formation mechanism for YMCs that contains
the normal mix of star formation processes which together produce
hierarchical structure and $M^{-2}dM$ power laws (see review in
Elmegreen et al. 2000). It applies to extensive galactic regions
with high pressures, such as merger remnants, ILR rings and
nuclear disks.

This research was supported by NSF grant AST-0205097 to BGE.

{}

\end{document}